\documentclass{scrartcl}

\usepackage{amsmath}
\usepackage[pdftex]{graphicx}
\usepackage{braket}

\newcommand{\Tr}{\mathrm{Tr}}

\title{A Matrix model from string field theory}
\date{}
\author{Syoji Zeze\footnote{ztaro21@gmail.com}\\ \\
Yokote Seiryo Gakuin High School,\\
147-1 Maeda, Osawa, Yokote, 013-0041 Japan}

\usepackage{graphicx}	% required for `\includegraphics' (yatex added)
\begin{document}

\maketitle

\begin{abstract}
We demonstrate that a Hermitian matrix model can be derived from level truncated open string field theory with Chan-Paton factors.  
The Hermitian matrix is coupled with a scalar and $U(N)$ vectors which are responsible for the D-brane at the tachyon vacuum.  Effective potential for the scalar is evaluated both for finite and large $N$.   Increase of potential height is observed in both cases.  The large $N$ matrix integral is identified with a system of $N$ ZZ branes and a ghost FZZT brane.

\end{abstract}

\section{Introduction}
Two different candidates for the nonperturbative formulation of string theory  have been known ---  
string field theory (SFT)  and matrix model.  It is commonly believed that they are just different descriptions of the underlying theory. Therefore, it is important to  investigate the relationship between these formulations. 
However, a few examples have been known to reduce open SFT (OSFT)~\cite{Witten:1985cc} 
to certain matrix models. 
First is OSFT for topological A or B model~\cite{hep-th/9207094} which reduces to Chern-Simons matrix
model~\cite{hep-th/0206255} 
or  ordinary Hermitian matrix model~\cite{hep-th/0207096}
respectively.  
Second is OSFT for (2,1) minimal string~\cite{hep-th/0312196} which 
reduces to Kontsevich matrix model~\cite{Kontsevich:1992ti}.  
In addition, less direct examples for $c=1$~\cite{hep-th/0503184},  $c=0$~\cite{hep-th/9307045} and critical~\cite{hep-th/9705128} strings have been known.
In those examples, each matrix model is obtained from different 
OSFT associated with particular boundary conformal field theory (BCFT).  
A systematic way to 
derive different matrix models from  OSFT in fixed background
has not yet known.  Finding such method
is important to study the background independence of OSFT. 

In this paper,  we present an example for such method.  
Our idea is simple: instead of varying BCFT,  we start
with a general setup in critical string, and approximate the 
string field to its first few components.   An example we will study is the \textit{level truncation} in the  
universal sector~\cite{hep-th/9911116} of critical ($D=26$) OSFT in which 
the string field is approximated at level $n$ as
\begin{equation}
 \Psi = \psi_{0} + \psi_{1} + \psi_{2} + \cdots + \psi_{n}.\footnote{Each component
 field carries Chan-Paton indices.}
\end{equation}
Given this approximation, the OSFT action~\cite{Witten:1985cc}
immediately reduces to a matrix action\footnote{
As we will see later, extra vectors and scalars couple
with the matrix.   Here we omit them for simplicity.
}
\begin{equation}
 S = a_{m n} \Tr \left[
 M_m M_n  \right]   +
 b_{m n p} \Tr   \left[ M_m M_n M_p\right], \label{141247_2Jul15} 
\end{equation}
where $M_n$ is a Hermitian matrix.   
The approximation is known to work well in the first few level and 
improves quickly as the level increases~\cite{hep-th/9912249, hep-th/0002237,
hep-th/0211012, arXiv:0910.3026, Kishimoto:2011zza}.

A possible reason that such method has not yet been 
examined is the lack of understanding of Chan-Paton factors in OSFT. 
Although Chan-Paton factors can be introduced to OSFT consistently, their origin has not yet been explained.  Recently,  
Erler and Maccaferri
proposed a new construction of classical
solutions of OSFT in terms of the 
regularized boundary conditions changing (BCC)
projectors~\cite{arxiv:1406.3021}. 
Their construction covers a wide range of backgrounds including 
multiple D-branes, which is the main interest of present paper.
Following their work, Kishimoto, Masuda, Takahashi and Takemoto (KMTT) 
demonstrated that $N$ Chan-Paton factors naturally arise from the
decomposition of string field in terms of the regularized 
BCC projectors~\cite{arxiv:1412.4855}. 
String field theory expanded around
the multiple D-branes solution
was interpreted as a system of $N+1$  D-branes.   
We will employ their formulation as our foundation.

This paper is organized as follows.  Section \ref{054043_3Jul15} 
introduces KMTT formulation.  In section \ref{012008_7Jul15}, 
we derive a one-matrix model in terms of  
the level truncation.    In section \ref{054237_3Jul15}, 
we derive the partition function 
of the model both in finite and large $N$.  Section \ref{054304_3Jul15} 
summarizes our results and further discussions are given.

\section{KMTT formulation}
\label{054043_3Jul15}

Let us briefly review the KMTT formulation~\cite{arxiv:1412.4855} 
of OSFT.  They begin with Erler-Maccaferri solution for
multiple D-branes~\cite{arxiv:1406.3021}  
\begin{equation}
 \Psi_0 = \Psi_{T} - \sum_{a=1}^{N} \Sigma_a \Psi_{T} 
 \bar{\Sigma}_a, \label{163417_20Jun15}
\end{equation}
where $\Psi_{T}$ is the tachyon vacuum solution and $\Sigma_{a}$ and
$\bar{\Sigma}_{a}$ are the regularized BCC projectors which obey
\begin{equation}
 \bar{\Sigma}_a \Sigma_b = \delta_{ab}, \qquad
  Q_{T} \Sigma_a = Q_{T}  \bar{\Sigma}_a =0.
\end{equation} 
Here $Q_{T}$ is the kinetic operator at the tachyon vacuum defined by
\begin{equation}
 Q_{T} \Psi  = Q_B \Psi + \Psi_{T} \Psi + \Psi \Psi_{T}.
\end{equation}
The index $a$ in $\Sigma_a$ or $\overline{\Sigma}_a$ is a Chan-Paton factor which labels D-branes.     
The solution \eqref{163417_20Jun15}  carries  degrees of freedom required
for multiple D-branes.  We further expect that  non-Abelian gauge gauge symmetry 
is realized in OSFT expanded around \eqref{163417_20Jun15}.  
KMTT~\cite{arxiv:1412.4855} introduced the decomposition of the string field which 
realizes this: 
\begin{equation}
 \Psi = \chi + 
 \chi_a \bar{\Sigma}_a + 
\Sigma_a \bar{\chi}_a 
+ \bar{\Sigma}_a \phi_{ab} \Sigma_b,
\label{eq:1}
\end{equation}
where $\chi$, $\chi_a$, $\bar{\chi}_a$ and $\chi_{a b}$ are component string fields.  
By expanding OSFT action around $\Psi_0$ according to \eqref{eq:1},  they
derived the matrix action
\begin{equation}
 S = S_1 + S_2 + S_3, \label{163204_21Jun15}
\end{equation}
\begin{equation}
 S_1 = -\frac{1}{g^2} 
\Tr \int  \left(
\frac{1}{2} \phi Q_{B} \phi  + \frac{1}{3} \phi^3
\right),\label{161644_21Jun15}
\end{equation}
\begin{equation}
 S_2 = -\frac{1}{g^2} 
\int  \left(
\frac{1}{2} \chi  Q_{T} \chi   + \frac{1}{3} \chi^3
\right),
\end{equation}
\begin{equation}
 S_3 = -\frac{1}{g^2} \int
\left(\frac{1}{2}    \bar{\chi_a} Q_{T0} \chi_a
+   \bar{\chi}_a  \chi  \chi_a
+  \bar{\chi}_a   \phi_{ab} \chi_b
\right),\label{162221_21Jun15}
\end{equation}
where $Q_{T0}$ is the kinetic operator defined by
$Q_{T0} \Psi  = Q_{B}\Psi + \Psi_{T} \Psi$. 
The trace in (\ref{161644_21Jun15}) runs over indices of $\phi_{a b}$,  and  identical indices in
(\ref{162221_21Jun15}) are to be summed over\footnote{%
We will ignore a constant shift in the action hereafter since it is not relevant 
for remaining discussions. 
}.
KMTT~\cite{arxiv:1412.4855} claimed that
 \eqref{163204_21Jun15} describes 
$N+1$ D-branes rather than $N$ D-branes.  
In addition to the  $N$ unstable D-branes, 
there is a ``D-brane'' at the tachyon vacuum described by $\chi$.
According to their interpretation, 
$\phi_{ab}$ connects two unstable D-branes $a$ and $b$ while
$\chi_a$ connects D-brane $a$ with the D-brane at the tachyon vacuum. 
$\chi$ represents fluctuation on the D-brane
at the tachyon vacuum. 

A remarkable feature of the action  (\ref{163204_21Jun15})
is the presence of the ``vector'' sector
described by $\chi_a$ and $ \bar{\chi}_a$, 
which has not been found in literature.  
The action $S_3$ is quadratic for the vectors therefore can be integrated out
in the path integral.   With assuming 
suitable gauge fixing, we perform path integral for $\chi_a$ and $\bar{\chi}_a$ 
and obtain a determinant factor
\begin{equation}
 \det(Q_{T0} + \phi + \chi)^{-1}\label{092653_3Jul15}
\end{equation}
in the partition function.   In the following sections, we will evaluate 
this determinant by truncating the string field rather than imposing
conventional gauge condition such as linear gauge \cite{Asano:2006hk}.   

\section{Truncation to matrices}
\label{012008_7Jul15}
Although the string fields presented in previous section carry
Chan-Paton factors,  they also depend on infinitely many labels
which distinguish each state in a BCFT. 
In order to make our analysis tractable, 
we  introduce an approximation in which
dynamical variables  reduce to matrices. 
Let us consider the string field in the universal sector\footnote{%
The open string field
in universal sector is build from the Virasoro generator $L_n$ and
conformal ghosts $b_n$ and $c_n$ on the $SL(2, R)$ vacuum~\cite{hep-th/9911116}.
} truncated at  level $n$,  where
the level is defined by eigenvalue of the kinetic operator of OSFT.
Denoting the level $k$ base  element of the string field $\psi_k$,
we truncate each component in (\ref{eq:1}) as
\begin{equation}
 \phi_{ab}   =  \sum_{k=1}^{n} M_{ab}^{(k)} \psi_k, \quad
 \chi_{a}    =  \sum_{k=1}^{n}  \xi_a^{(k)} \psi_{k}, \quad
 \bar{\chi}_{a}    =  \sum_{k=1}^{n}  \bar{\xi}_a^{(k)} \psi_{k}, \quad
 \chi  =  \sum_{k=1}^{n}  t^{(k)}    \psi_{k},\label{160118_5Jan15}
\end{equation}
where $M_{ab}^{(k)}$ is a Hermite matrix, $\xi_{a}^{(k)}$ is a complex vector,   $\bar{\xi}_{a}^{(k)}$ is its complex conjugate, and $t^{(k)}$ 
is a real number.
Note that these component fields do not depend on 
any other variables.   As a concrete example of such truncation, we choose
the expansion examined in Ref.~\cite{arxiv:1004.4351} for
dressed $\mathcal{B}_0$ gauge~\cite{arxiv:0906.0979}.     In this case, the level $k$ base is given by
\begin{equation}
 \psi_k = c  K^n  B c \frac{1}{1+K},\label{070927_3Jul15}
\end{equation}
where $K, B, c$  are the elements of the $K B c$  algebra~\cite{hep-th/0603159}.    We would like to study level 1 truncation in which string fields are
given by
\begin{align}
 \phi_{a b} & = M^{(0)}_{a b} \psi_{0}+ M^{(1)}_{a b} \psi_{1}, \\
 \chi_{a} & = \xi^{(0)}_{a } \psi_{0}+ \xi^{(1)}_{a } \psi_{1}, \\
 \bar{\chi}_{a} & = \bar{\xi}^{(0)}_{a } \psi_{0}+ \bar{\xi}^{(1)}_{a} \psi_{1}, \\
  \chi & = t^{(0)} \psi_{0}+ t^{(1)} \psi_{1}.
\end{align}
A matrix action obtained by this truncation is rather complex since it includes both level 0 and 1 fields.  However,  it is possible to reduce degrees of freedom further.  It is know that  level 1 fields cannot have cubic terms since \cite{arxiv:1004.4351}
\begin{equation}
\int \psi_{1}^{3} =0.
\end{equation}
Thus, level 1 fields are quadratic in the action and can be integrated out using equations of motion.   With the help of the explicit values of products
of the basis $\psi_0$ and $\psi_1$ evaluated in \cite{arxiv:1004.4351}, a nontrivial solution of equations of motion for $M^{(1)}_{ab}, \xi^{(1)}_{a}, \bar{\xi}^{(1)}_a$ and $t^{(1)}$  turns out to be
\begin{equation}
 M_{a b}^{(0)} =  M_{a b}^{(1)},  \quad
 \xi_{a}^{(0)} =  \xi_{a}^{(1)},  \quad
 \bar{\xi}_{a}^{(0)} =  \bar{\xi}_{a}^{(1)},  \quad
 t^{(0)} =  t^{(1)}.
\end{equation}
By this assignment,  four string fields  in \eqref{160118_5Jan15} become
proportional to Erler-Schnabl solution~\cite{arxiv:0906.0979}:
\begin{equation}
 \Psi_{T}  = c (1+K) B c \frac{1}{1+K}.
\end{equation}
Thus, the four fields\footnote{%
This truncation makes sense if we replace $\Psi_T$ 
with another representation of the tachyon vacuum solution, since
we do not require its explicit form in following analysis.
}
can be written as
\begin{equation}
 \phi_{ab}   =  M_{ab} \Psi_T, \quad
 \chi_{a}    =  \xi_a \Psi_{T}, \quad
 \bar{\chi}_{a}    =  \bar{\xi}_a \Psi_{T}, \quad
 \chi  =    t   \Psi_{T}. 
\end{equation} Although only few fields are included in the approximation, 
 we expect that it well
captures the essence of the dynamics of D-branes,
as is the case in the conventional level truncation analysis.  
An evidence that supports our expectation is that
the truncation interpolates between two analytic solutions:
one is the Erler-Maccaferri's multiple D-branes denoted as
$(M_{ab}, \xi_a, \bar{\xi}_a, t) = (0,0,0,0)$ while the other is the 
perturbative vacuum, also denoted as  $(\delta_{ab},0,0,-1)$.

Let us derive the truncated action.  The requirement that $\Psi_{T}$ reproduces the correct 
value of the D-brane tension is represented by
\begin{equation}
 \int \Psi_{T} Q_B \Psi_{T} = - \frac{3}{\pi^2}, \quad
  \int \Psi_{T} Q_{T0} \Psi_{T}= 0, \quad
 \int  \Psi_{T} Q_{T} \Psi_{T} = +\frac{3}{\pi^2}, \quad
\quad  \int \Psi_T^3 = + \frac{3}{\pi^2}.
\end{equation}
Applying these to  (\ref{163204_21Jun15}),
we find
\begin{equation}
 S  = -\frac{1}{g'^2}  \left[ \Tr \left(
-\frac{1}{2} M^2 + \frac{1}{3} M^3  \right)
+ \frac{1}{2 } t^2 + \frac{1}{3 } t^3 
+ \bar{\xi}_a (t \delta_{a b} +M_{a b} ) \xi_b 
\right],\label{152154_24Dec14}
\end{equation}
where we have  rescaled  the open string coupling as
\begin{equation}
 \frac{1}{g'^2} = \frac{3}{g^2 \pi^2}. 
\end{equation}
For later convenience, we shift
$M$ to $M+1$ and omit the prime in $g'$. 
 Then, ignoring constant shift, we obtain an action
\begin{equation}
  S  = -\frac{1}{g^2}  \left[ \Tr\,  W(M)
+ W(t) +\bar{\xi} ( 1+t +M ) \xi 
\right],\label{220347_24Jun15}
\end{equation}
where $W (x) = \frac{1}{2} x^2  + \frac{1}{3} x^3$.  In this way, the truncation
(\ref{160118_5Jan15}) reduces OSFT action to a cubic matrix action
coupled with a scalar and  complex vectors\footnote{%
A matrix model with $U(N)$ vector was proposed in
\cite{hep-th/0305159} in a different setting.}.   
The partition function  for (\ref{220347_24Jun15})
can be obtained by employing the standard technique of matrix model\footnote{
For example, see \cite{hep-th/0410165}.}.
As readily found in (\ref{220347_24Jun15}), the action is invariant
under $U(N)$ transformation 
 $M\rightarrow U M U^\dagger,\  \xi \rightarrow U \xi,\ \bar{\xi}
\rightarrow \bar{\xi} U^{\dagger}$, where $U$ denotes
an $U(N)$ matrix.    This is a residual gauge symmetry of  KMTT action \cite{Kishimoto:2014yea}.   
This $U(N)$ symmetry can be fixed by
diagonalizing $M_{ab}$ to its eigenvalues $\lambda_a$ with inserting
Vandermonde determinant $(\lambda_a -\lambda_b)^2$ in the partition
function.   After performing Gaussian integral for $\xi$, 
we obtain a partition function
\begin{equation}
 Z = \int  \prod_{a=1}^{N} d\lambda_{a} d t\, 
 e^{- V },\label{040553_6Jan15}
\end{equation}
where
\begin{equation}
 V  =  \frac{1}{g^2}
\left(  W(t) + \sum_{a=1}^{N} W(\lambda_a)  \right)
+  \sum_{a=1}^{N}   \log    |t+\lambda_a + 1|
-  \sum_{a< b} \log(\lambda_a -\lambda_b)^2. \label{173139_5Jan15}
\end{equation}
This can describes  a system of
$N+1$ particles moving in the potential $W$. 
An eigenvalue $\lambda_a$ feels repulsive forces from other eigenvalues 
through the last term of \eqref{173139_5Jan15}.   Also, 
it is  attracted towards $-t-1$ due to the second term in \eqref{173139_5Jan15}. 

It is interesting to compare our result with the matrix 
formulation of $c < 1$ noncritical string theories.  It has been 
recognized that Hermitian one-matrix model serves
nonperturbative definition of $(2, 2 k +1)$ minimal string theory. 
As an example, let us consider the action studied in Ref.~\cite{hep-th/0406030}:
\begin{equation}
 S  = -\frac{1}{g^2}  \left[ \Tr\,  W(M) + \bar{\psi} (M -z) \psi 
\right],\label{220034_25Jun15}
\end{equation}
where $\psi$ and $\bar{\psi}$ are   \textit{fermionic} vectors. They can be integrated out so that insert a factor
\begin{equation}
 \det (M-z)\label{060544_12Jul15}
\end{equation}
in the matrix integral.  In double scaling limit, 
this determinant is identified with a FZZT brane~\cite{hep-th/0001012,
hep-th/0009138}, and the fermionic vectors
are identified with fermionic strings between a FZZT brane and a stuck of
$N$ ZZ branes~\cite{hep-th/0101152}.   In contrast, our action 
(\ref{220347_24Jun15}) contains \textit{bosonic} vectors $\xi$  and $\bar{\xi}$
rather than fermionic ones.   Integration with respect to them yields a factor
\begin{equation}
{\det (M+t+1)}^{-1}\label{095328_6Jul15}
\end{equation}
in the partition function. Comparison between (\ref{060544_12Jul15})
and (\ref{095328_6Jul15}) naturally identifies 
the determinant (\ref{095328_6Jul15}) 
as a \textit{ghost} FZZT brane \cite{hep-th/0601024}
which cancels the effect of a FZZT brane
\footnote{The ghost D-brane has been proposed as an object which cancels the effects of a D-brane  \cite{hep-th/0601024} . }.  
Unfortunately, corresponding observable in minimal string theory is
not yet  identified.

\section{Effective potential}
\label{054237_3Jul15}

In this section, we will evaluate the effective potential for
$t$ in terms of saddle point equations for eigenvalues.  
In large $N$ limit of t'Hooft expansion, 
saddle point configurations are  leading contributions 
to the matrix integral.  
Even for finite $N$, saddle point configurations also offer
a good approximation to the matrix integral when $g$ is small.  
In either case, saddle point equations are obtained from
the variation  of (\ref{173139_5Jan15}):  
\begin{gather}
 \frac{1}{g^2}   W'(\lambda_a) %( \lambda_a +\lambda_{a}^{2} )
+ \frac{1}{t + \lambda_a + 1} -2 
\sum_{b \neq a}^{N} \frac{1}{\lambda_a-\lambda_b}  =0, \quad (a = 1 \dots N)  ,\label{174450_5Jan15}\\
 \frac{1}{g^2} W'(t)
%(t +t^2 ) 
+\sum_{a=1}^{N} \frac{1}{t+\lambda_a +1}  = 0.\label{174502_5Jan15}
\end{gather}

\subsection{$N=1$}

\begin{figure}[htbp]
\centering
 \includegraphics[scale=.5]{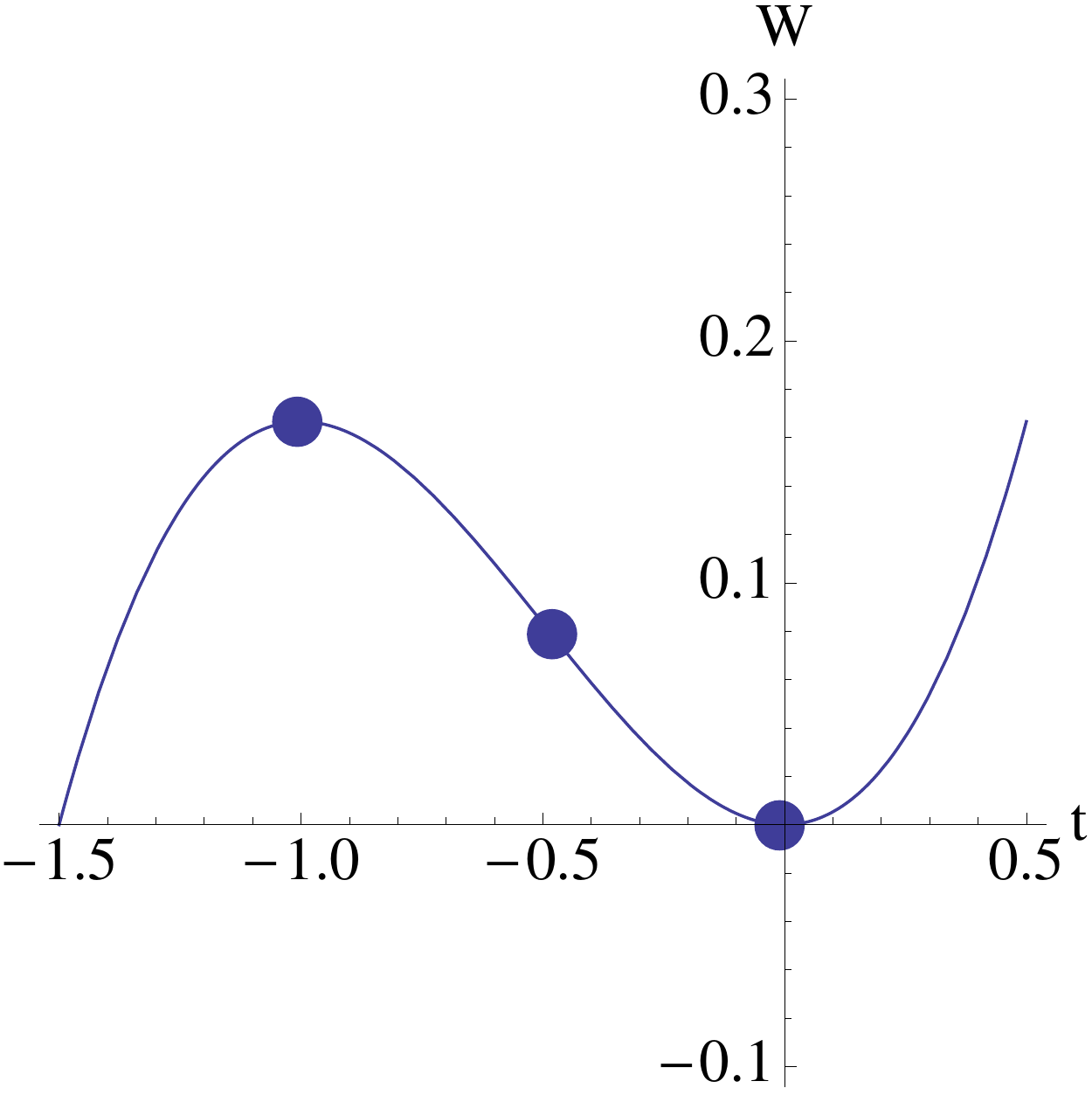}
\caption{Solutions at $g=0.10$}
\label{111118_27Jun15}
\end{figure}

\begin{figure}[htbp]
\centering
 \includegraphics[scale=.5]{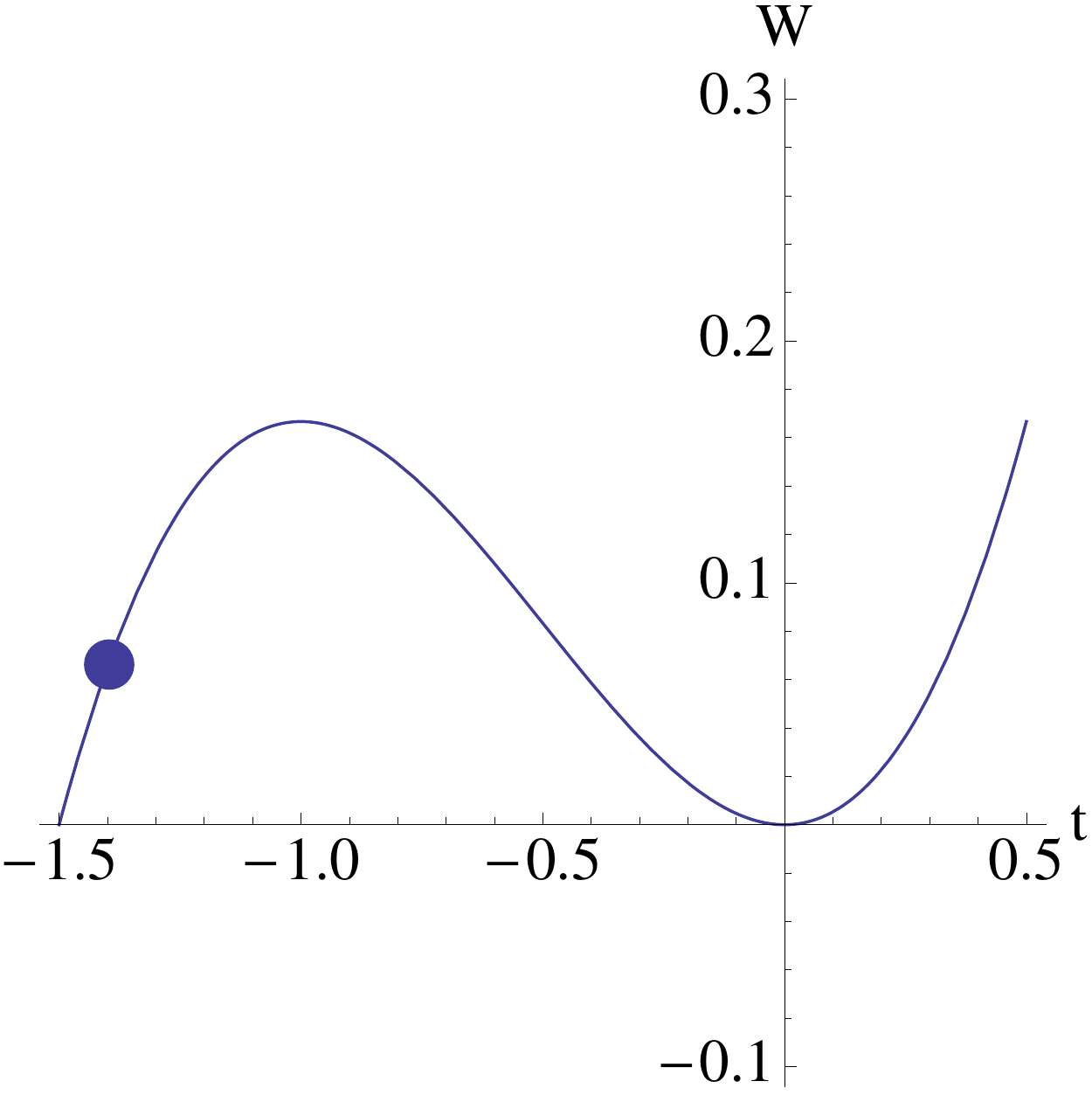}
\caption{Solution at $g=1.0$.}
\label{111144_27Jun15}
\end{figure}

Let us study the dynamics of the model at finite $N$.  We begin with
$N=1$, which  corresponds to a 
system of an unstable D-brane and another D-brane
at the tachyon vacuum.  
Denoting $\lambda_a$ as $\lambda$, 
the potential (\ref{173139_5Jan15}) is 
given by
\begin{equation}
V  =   
 \frac{1}{g^2} \left(
W(t) + W(\lambda )
\right)  +    \log  |t+\lambda+1|,  \label{190909_5Jan15}
\end{equation}
and saddle point equations are
\begin{gather}
\frac{1}{g^2}
(\lambda+\lambda^2) +  \frac{1}{t+\lambda+1}  = 0,\label{110209_25Dec14} \\
\frac{1}{g^2}
(t+t^2) + \frac{1}{t+\lambda+1 }  =0.\label{110219_25Dec14}
\end{gather}
By combining above two equations, we obtain a coupling independent
equation
\begin{equation}
 t(t+1) = \lambda (\lambda+1),
\end{equation}
which has two roots $\lambda = t$ and $\lambda = -t -1$.  
The  latter is not appropriate since it hits the singularity
$(t+\lambda +1)^{-1}$ in the partition function.  
Therefore, we choose $t=\lambda$ as our solution. 
  Given this choice, (\ref{110209_25Dec14})
and (\ref{110219_25Dec14})  reduce to single equation
\begin{equation}
\frac{1}{g^2}
(t + t^2 ) +  \frac{1}{2 t +1}  = 0.
\end{equation} 
This can be rewritten to a cubic equation,
\begin{equation}
t (t+1) (2 t +1) + g^2 = 0
\end{equation}
 whose  roots 
correspond to  saddle points.   
It is easily understood that while there are three roots 
for small $g$, two of them disappear beyond critical value of $g$. 
Let us discuss further details as follows. 
Small $g$ expansion of these roots is
\begin{align}
-g^2 -3 g^4 +\mathcal{O} (g^6), \quad 
-1 -g^2 + 3 g^4 +\mathcal{O} (g^6), \quad
-\frac{1}{2} + 2 g^2 + \mathcal{O} (g^6).
\end{align}
It is useful to show positions of these roots in a plot of  the potential
$W$.   Fig.~\ref{111118_27Jun15}
and \ref{111144_27Jun15} are such plots for different values of $g$.  Fig.~\ref{111118_27Jun15} shows that saddle points for small $g$; 
 saddle points are placed within the region $(-1, 0)$ with equal intervals.
As $g$ become larger, two larger roots 
become closer and annihilate beyond the critical value
\begin{equation}
  g_{c} = 2^{-1/2} 3^{-3/4} \simeq 0.31.
\end{equation}
Fig. \ref{111144_27Jun15} shows a placement of a root after the
annihilation.   

Analysis made here is merely a classical approximation and 
not full quantum treatment. 
Fortunately,  we can integrate out one variable in the potential 
without relying on the saddle 
point approximation since present example is enough simple.
Let us change the variables as 
\begin{equation}
 u =\frac{t+\lambda}{2},\qquad v =\frac{t-\lambda}{2}.
\end{equation}
Then, the partition function can be written as
\begin{align}
 Z & = \int d u d v 
\frac{1}{|2 u + 1|} \exp \left\{ 
\frac{1}{g^2 } \left(- 2 W(u) - (2 u + 1) v^2
 \right)
\right\}\label{071139_3May15} \\
  & = 
 \int d u \frac{1}{|2 u + 1|^\frac{3}{2}}
\exp \left( -\frac{2}{g^2} W(u) \right)
\end{align}
where we performed Gaussian integral for $v$ in the second line.  Thus,
we obtain one dimensional effective potential
\begin{equation}
 V_{\mathrm{eff}} (u) = \frac{2}{g^2} W(u)
+ \frac{3}{2} \log |2 u +1|.\label{121209_27Jun15}
\end{equation}
We note that the saddle point approximation corresponds to
setting $v=0$ in (\ref{071139_3May15}), since we our choice of the saddle point,   $\lambda = t$
corresponds to it.   Therefore, 
\begin{equation}
  V_{\mathrm{saddle}} (u) = \frac{2}{g^2} W(u)
+ \frac{1}{2} \log |2 u +1|. \label{nointegeq}
\end{equation}
By comparing \eqref{121209_27Jun15} and \eqref{nointegeq}, we find that they only differ in the coefficient of the logarithmic term which is negligible for small $g$.  Thus, 
the saddle point approximation works well for small coupling.

\begin{figure}[htbp]
\centering
 \includegraphics[scale=.7]{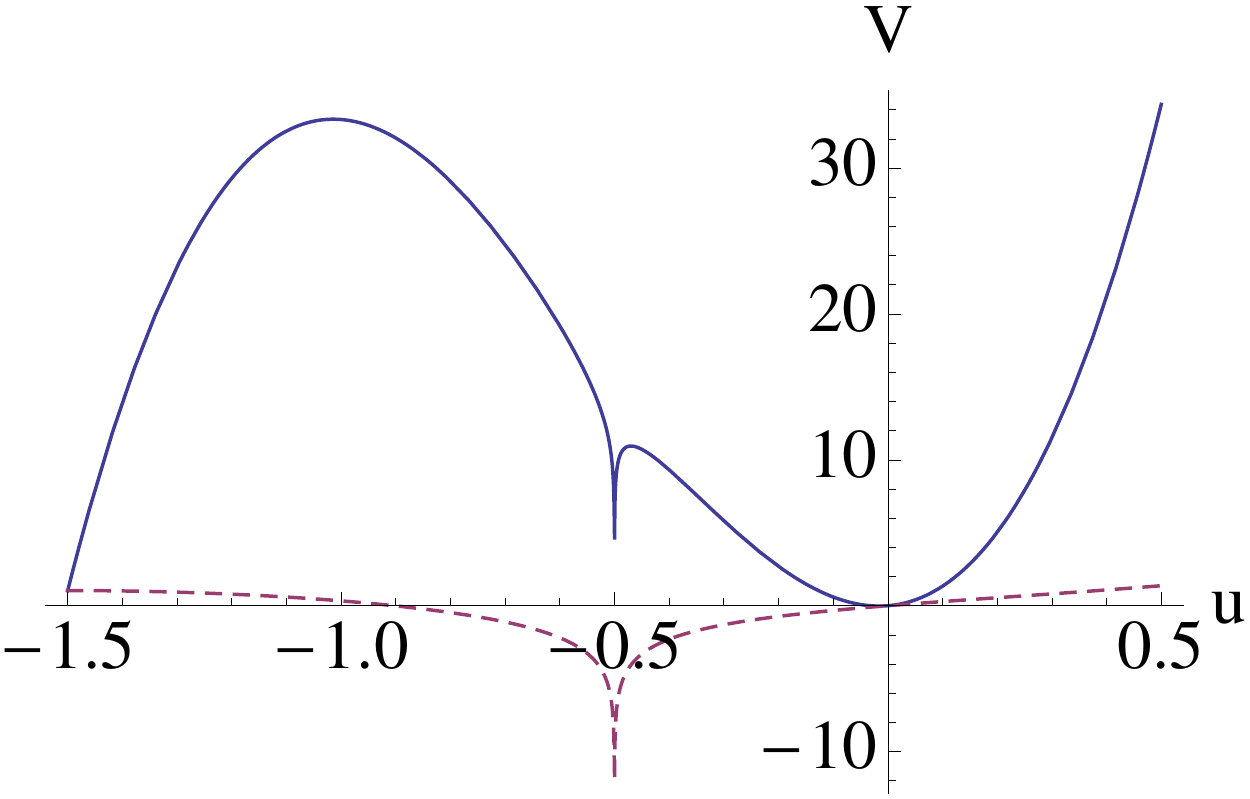}
\caption{Effective potential for $u$, where solid line 
is a plot for $g=0.1$ while dashed line is for $g=1.0$. }
\label{121145_27Jun15}
\end{figure}

Finally, let us compare the shape of the potential for different values
of $g$.   Fig. \ref{121145_27Jun15} is a plot of (\ref{121209_27Jun15})
for small and large $g$.   It is observed that the stable vacuum 
around $u=0$ disappears for large $g$ due to dominance of the logarithmic term.  
This phenomenon corresponds to the annihilation of saddle points which has already
been observed in the saddle point analysis.   Such dependence of the
effective potential on the coupling is consistent with the fact that
 the D-brane describes the system at small coupling.     

\subsection{$N \geq 2 $}

Next we proceed to $N\geq 2$, where the equations of motions are given 
by (\ref{174450_5Jan15}) and (\ref{174502_5Jan15}).  
We again perform  saddle point approximation by finding roots of these equations.  
The roots can be obtained numerically for each value of $g$.  
Numerical solutions for $g=0.01$ and up to $N=3$ are shown in tables
\ref{074633_6May15}, \ref{074647_6May15} and \ref{074656_6May15}.   
We  pick real roots only, and each
solution is specified as $(t, \lambda_1, \dots, \lambda_N)$.  Due to the
symmetry which exchanges $\lambda$s, we only need to specify single 
configuration for each value of $t$ among $N$ permutations of $\lambda.$
We also evaluate the value of effective potential $V(\lambda, t)$
normalized by the value of D-brane tension $1/(6 g^2)$. 
\begin{table}[htbp]
\centering
\begin{tabular}{ccc} \hline\hline
$t$ & $ \lambda$ &  $V$ \\  \hline
$-1.0001$ & $-1.0001$   & 2.0000            \\
$-0.4799$ & $-0.4799$  & 0.9947            \\
$-0.0001$ & $-0.0001$  & 0.0000 \\
\hline\hline
\end{tabular}
\caption{Numerical solutions for $N=1$ at  $g=0.01$. }
\label{074633_6May15}
\end{table}
\begin{table}[htbp]
\centering
\begin{tabular}{cccccc} \hline\hline
$t$ &$ \lambda_1 $ & $ \lambda_2 $&
$\lambda$ permutations  &  $V$  \\  \hline
$-0.6664$ &$ -0.9998 $ &  $ -0.3330$ & 2   & 1.9950         \\
$-0.3331$ & $-0.6664$ & $ 0.0098$ & 2  & 0.9950            \\
$-0.0002$ & $ 0.0098, $ & $ -0.0102$  & 2    & 0.0052   \\
\hline\hline
\end{tabular}
\caption{Numerical solutions for $N=2$ at  $g=0.01$. }
\label{074647_6May15}
\end{table}
\begin{table}[htbp]
\centering
\begin{tabular}{ccccccc} \hline\hline
$t$ &$\lambda_1 $  & $\lambda_2 $ & $\lambda_3 $
& $\lambda$ permutations &  $V$ \\  \hline
$-0.4997$  & $-0.9996$  & $-0.4999$   &$-0.0003$ & 6   & 1.9955         \\
$-0.2498$ & $-0.7496$  & $-0.0100$& $-0.0099$ &   6   & 1.0005            \\
$-0.0003$& $-0.0176$  & $-0.0169$   & $-0.0002$   & 6      & 0.0155           \\
\hline\hline
\end{tabular}
\caption{Numerical solutions for $N=3$ at  $g=0.01$.}
\label{074656_6May15}
\end{table}
We observe an interesting pattern in these results.
Each root is located at one of a 
`site' which is obtained by dividing the region $-1 \leq \lambda \leq  0
$ into $N+1$ intervals, i.e.,    
\begin{align}
 x_{m} = - \frac{m}{N+1},  \quad (m=0,1, \dots N+1).
\end{align}
We also observe that the minimum value of $t$ is given by $-2/(N+1)$ while
values of $\lambda$s can reach $-1$.   We also observed that the sum of
all eigenvalues are close to its value of potential, i.e., 
\begin{equation}
 t +\sum_{a=1}^{N}\lambda_a \sim V.
\end{equation}
Finally, it is also interesting to see that the maximum value 
of the potential is close to $2$ for all $N$.  

We would like to close this subsection with a summary of our result:
 \begin{itemize}
 \item  The stable vacuum is lost for large coupling.  This indicates
	breakdown of D-brane description.   The critical value of the
	coupling can be determined by saddle point equations.  
  \item At small $g$, the local maximum of the effective potential 
        is about twice higher than that of original one.
\end{itemize}

\subsection{Large $N$}
Let us evaluate the partition function in large $N$ limit
following with the standard method of matrix model~\cite{hep-th/0410165}.  
We begin with  the partition function
\begin{equation}
 Z  = % \frac{ (2 \pi)^N}{N!} 
\int d\lambda_{_a} d t \frac{\prod_{a < b} (\lambda_a-\lambda_b)^2  }{\prod_{a}
 |t+\lambda_a +1|}  e^{-\frac{1}{g^2}  \left(
  W(t)  +   \sum_{a=1}^{N} W (\lambda_a )    \right)}
\end{equation}
and introduce the t'Hooft coupling
\begin{equation}
 \mu = g^2 N.  
\end{equation}
Then, saddle point equations read
\begin{equation}
\frac{N}{\mu} (t+t^2) +  \sum_{a} \frac{1}{\lambda_a + t +1} =0,\\
\end{equation}
\begin{equation}
\frac{N}{\mu} (\lambda_a+\lambda_a^2) + 
  \frac{1}{\lambda_a + t +1} 
- \sum_{b} \frac{2}{\lambda_a -\lambda_b} 
=0. \\ 
\end{equation}
In large $N$ limit, eigenvalues are described by
a continuous distribution $\rho (\lambda)$ and  summation 
for eigenvalues is replaced with  integration  
\begin{equation}
 \frac{1}{N} \sum_{a} \rightarrow \int d\lambda \rho (\lambda).
\end{equation}
Then, saddle point equations  are replaced with
\begin{equation}
 \frac{1}{\mu} (t + t^2)  +   \int d\lambda 
\frac{\rho (\lambda )}{t+\lambda +1} =0,
\end{equation}
\begin{equation}
 \frac{1}{\mu} (\lambda + \lambda^2)  + 
\frac{1}{N} \frac{1}{t+\lambda +1}
- 2 
\int d\lambda' 
\frac{\rho (\lambda' )}{\lambda -\lambda'} =0.\label{072108_7Jul15}
\end{equation}
At leading order, the second term in (\ref{072108_7Jul15})
becomes negligible.   Thus we obtain saddle point equations 
\begin{equation}
 \frac{1}{\mu} (t + t^2) +  \int d\lambda 
\frac{\rho (\lambda )}{\lambda + t +1} =0,\label{172740_5Jun15}
\end{equation}
\begin{equation}
 \frac{1}{\mu} (\lambda + \lambda^2)  
- 2 
\int d\lambda' 
\frac{\rho (\lambda' )}{\lambda -\lambda'} =0.
\label{lleq}
\end{equation}
The latter equation \eqref{lleq} solves the planar limit of a cubic matrix model
whose solution can be found elsewhere \cite{Brezin:1977sv}. 
It is convenient to introduce the resolvent 
\begin{equation}
 \omega (z)  = \int d \lambda \frac{\rho (\lambda)}{\lambda-z}.
\end{equation}
The equation \eqref{lleq} can be replaced with an equation for $\omega(z)$.  Once $\omega (z)$ is obtained, \eqref{172740_5Jun15} can be solved by finding solutions of
\begin{equation}
 \frac{1}{\mu} (t+t^2) + \omega (-t -1) =0.
\end{equation}
The ``one-cut'' solution for the resolvent in large $N$ limit 
is known to be \cite{Brezin:1977sv} 
 \begin{equation}
 \omega (z) = \frac{1}{2 \mu} \left\{
z+ z^2 - \sqrt{(z-a)(z-b)}\left(1 +\frac{a+b}{2} + z \right)
\right\},\label{132753_27Jun15}
\end{equation}
where $a$ and $b$ are endpoints of the branch cut, which 
define a support for the eigenvalue density $\rho (\lambda)$.  Requiring 
$\omega (z) \sim z^{-1}$ at infinity, 
we obtain equations 
\begin{equation}
 3(a+b)^2 + 4 (a+b) -4 a b =0, \label{152002_27Jun15} 
\end{equation}
\begin{equation}
 (a+b)^3 + (a+b)^2 -4 a b (a+ b +1) =16 \mu. 
\end{equation}
which solve $a$ and $b$ as functions of $\mu$. 
 These equations are conveniently rewritten
in terms of parameters $\sigma = a + b$ and $\bar{\sigma}=a-b$ as
\begin{equation}
 16 \mu = - \sigma (\sigma +1) (\sigma+2),\label{141727_27Jun15}
\end{equation}
\begin{equation}
 2 (\sigma +1)^2 +\bar{\sigma} =2.
\end{equation}
The latter equation restricts $\sigma$ inside 
$-2 \leq \sigma \leq 0$.  Further,
(\ref{141727_27Jun15}) tells us that there are no real 
roots of (\ref{141727_27Jun15})
within $-2 < \sigma < 1$.  Therefore $\sigma$ is constrained within
\begin{equation}
 -1 \leq \sigma \leq 0.
\end{equation}
Let us choose a brunch which starts from $\sigma=0$.  From (\ref{141727_27Jun15}), 
the maximum value of $\mu$ reads
\begin{equation}
 \sigma_{c} = \frac{1}{3} (-3 + \sqrt{3}) \sim -0.42, \qquad
 \mu_{c} =\frac{1}{24 \sqrt{3}} \sim 0.024
\end{equation}

Given these ingredients, equation (\ref{172740_5Jun15}) can be rewritten into
\begin{equation}
\frac{1}{\mu} \left\{
\frac{3}{2} (t+t^2) + \frac{1}{2}\sqrt{t^2 + (\sigma +2 ) t +
\frac{3}{4} \sigma^2 + 2 \sigma +1    }\left(t- \frac{\sigma}{2}\right)
\right\} = 0.
\end{equation}
Integrating this equation yields an effective potential for $t$,
\begin{equation}
 V (t) = \frac{1}{\mu}( v_1 (t)  + v_2 (t) ) \label{effpotlargeN}
\end{equation}
where
\begin{equation}
v_1 (t) =  \frac{3}{2} W(t),
\end{equation}
\begin{equation}
v_2 (t) = \frac{1}{2}  \int_{0}^{t} dz\, 
\left(z- \frac{\sigma}{2}\right) 
\mathrm{Re} \sqrt{z^2 + (\sigma +2 ) z + \frac{3}{4} \sigma^2 + 2 \sigma
+1    }.
\end{equation}
\begin{figure}[htbp]
\centering
 \includegraphics{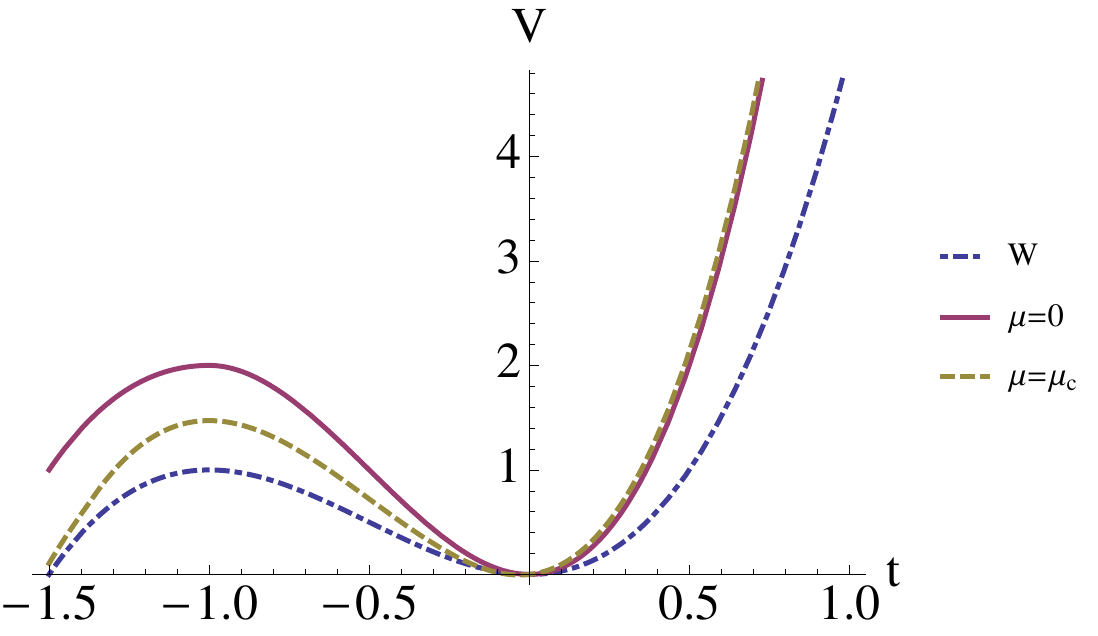}
\caption{Effective potential at large $N$. The value of the potential
is normalized by $1/(6 \mu)$.  The original potential $W(t)$ is
shown together.}
\label{070117_3Jul15}
\end{figure}
A plot of effective potential \eqref{effpotlargeN} is shown in Fig.~\ref{070117_3Jul15} .  
The local minimum at $t=0$ and the local maximum around $t=-1$ are observed 
for any value of $\mu$.   
This is consistent with the t'Hooft limit, 
in which $g$ is small so that the D-brane description of the system
holds.  On the other hand, 
the height of the local maximum is always higher
than that of the original potential $W$.   
The height is maximum for $\mu = 0$; in this case, the potential  is twice higher than 
$W$.   While the height decreases
as $\mu$  increases, it remains higher than $W$ even at a maximum value of $\mu=\mu_c$.
The increase of the height from the original one 
can be understood from the particle description 
of eigenvalues.  Eigenvalues filled in the bottom of $W(t)$
pull the tachyon $t$ by attractive force.  It makes harder for $t$
to climb the potential wall, thus increases  the height of the
potential height.    Such dependence of the effective potential on $\mu$ 
is quite different 
from that of the probe eigenvalue model 
investigated in \cite{hep-th/0405076} where a
plateau along the eigenvalue distribution is observed.   The 
existence of the plateau is explained by the fact 
that a probe eigenvalue cannot be distinguished from others.
Our model has no plateau since $t$ can be distinguished from other
eigenvalues.

\section{Conclusions and discussions}
\label{054304_3Jul15}

In this paper,  we proposed a systematic method to derive
matrix models from level truncated  OSFT.  Obtained 
matrix model contains $U(N)$ vectors and a scalar in addition
to Hermite matrix.  We have evaluated the effective potential of
the scalar both for finite and large $N$.  Increase of 
the potential height was observed at small 
coupling.   In section \ref{012008_7Jul15}, 
we have interpreted our model as a system of 
ZZ branes and a ghost FZZT brane. 

We would like to discuss further issues to be explored.  First, 
we would like to present an alternative but  rather heuristic
interpretation of our result.
Let us go back to the inverse of the determinant : 
\begin{equation}
 \det (Q_{T 0} + \phi +\chi)^{-1}.\label{232312_6Jul15}
\end{equation}
The basic idea is that this quantity 
can be regarded as a propagator with 
$\phi$ and $\chi$ insertions. Recall that this factor is
obtained by an integrating out  $\chi_a$  and $\bar{\chi}_a$ which connect
a D-brane with the tachyon vacuum where the world-sheet boundary disappears.
Therefore, it is natural to think that (\ref{232312_6Jul15}) 
amounts to a disk amplitude with single world-sheet boundary of a D-brane.  
Schematically, such disk amplitude can be written
\begin{equation}
 - \Tr \int \frac{d t}{t} e^{- t (Q_{T 0} + \phi +\chi) }.
\end{equation}
As is well known,
small $t$ limit of such amplitude corresponds to 
closed string propagation~\cite{hep-th/9510017}.  Therefore,
the inverse determinant \eqref{232312_6Jul15} encodes gravitational
force between $N$ D-branes and the tachyon vacuum\footnote{%
This interpretation is consistent with the attractive force
between $\lambda$ and $t$ observed in section \ref{012008_7Jul15}. 
}.   

Second issue is abut the level truncation.
We have seen that  the approximated OSFT action at
first few levels yields a one-matrix model which can be 
interpreted as $c<1$ noncritical string theory~\cite{hep-th/0408039}.
It is interesting to improve the approximation by 
including higher level fields to obtain
multi-matrix models.   We speculate that
that the improved matrix action
continues to be dual to some closed string theory on more nontrivial background.
Finally, we will recover original OSFT  with 
infinitely many matrices in infinite level limit.   We also expect that this
matrix model describes a critical string theory in nontrivial background 
through AdS/CFT like duality \cite{Maldacena:1997re}.   Thus, the  level truncated OSFT offers a way to describe closed strings in approximated geometry.   

Last issue is the matrix description
of  OSFT based on the left-right splitting of open strings which have been examined in past \cite{%
hep-th/0105058,hep-th/0105059, hep-th/0106036, hep-th/0107101}.  
Our model based on the KMTT decomposition 
looks quite differently from these models.  However, as mentioned in the previous paragraph, our model will recover full OSFT  in infinite level.  Thus the left-right splitting models and our model both describe same OSFT.    
We expect that all ingredients of KMTT, including Chan-Paton factors and 
regularized BCC projectors, are embedded in the left-right type matrix model 
in quite nontrivial manner.

Together with recent development which deals with 
different backgrounds as classical solutions~\cite{arxiv:1406.3021}, 
our result presents further evidence for SFT as a formulation of 
nonperturbative string theory.   We hope that further developments in 
this direction will shed light on the landscape of string theory.

\end{document}